# Data-driven design of new catalytic materials in methane oxidation based on a site isolation concept

A. Mazheika[1,*], M. Geske[1], M. Müller[2], S.A. Schunk[2,3,4], F. Rosowski[1,3], R. Kraehnert[1]

[1]BasCat (UniCat BASF JointLab), Technische Universität Berlin, Berlin, Germany

[2]hte GmbH, Heidelberg, Germany

[3]BASF SE, Catalysis Research, Ludwigshafen, Germany

[4]Institute for Chemical Technology, University of Leipzig, Germany

*corresponding author: mazheika@tu-berlin.de, alex.mazheika@gmail.com

## Abstract

The conversion of natural gas (methane) to ethane and ethylene (OCM: oxidative coupling of methane) facilitates its transportation and provides a way to synthesize higher value chemicals. The search for high-performance catalysts to achieve this conversion is the main scope of most corresponding studies in the field of OCM. Here, we present a general data-driven strategy for the search of novel catalytic materials, focusing particularly on materials useful for the OCM reaction. Our strategy is based on consistent experimental measurements and includes *ab initio* thermodynamics calculations and *active* screening. Based on our experiments, which showed unique volcano-type dependence of the performance on the stability of formed carbonates attributed to the site isolation concept, we developed a method for efficient and inexpensive DFT calculations of the formation energies of carbonates with prediction accuracy 0.2 eV. This method was implemented into a high-throughput screening scheme, which includes both general requirements for catalyst candidates and an actively done artificial intelligence part. Experimental validation of some of the candidates obtained during the screening showed successful reproduction of the initial volcano dependence. Moreover, several new materials were found to outperform standard OCM catalysts, specifically at lower temperatures.

# Introduction

Methane is the main component of natural gas, contributing up to 90% in volume of its content. The $CH_4$ molecule itself is a rather stable, closed-shell molecule. This means its conversion is a more challenging task than for higher hydrocarbons, in which C-H bonds are weaker due to inductive effects. The OCM is a direct way to convert methane into higher hydrocarbons, especially ethane ($C_2H_6$) and ethylene ($C_2H_4$) [1]. The conversion to $C_2$ products is the entry port for further synthesis of many chemicals from natural gas, including polymers, plastics, etc. The OCM reaction is triggered by oxides under rather high temperatures – above 600°C in an oxygen-rich atmosphere – and includes both heterogeneous and homogeneous steps [2,3]. Despite decades of intensive research, however, no economically viable catalyst [4,5] has been reported yet. The main issue is that under high reaction temperatures, methane and a number of the primary products are easily combusted to carbon oxides (low selectivity). On the other hand, lower reaction temperatures (below 600°C) are in general not high enough to overcome kinetic barriers [6] which lead to low conversion and consequently low yields per pass. As a result, an ideal OCM catalyst must combine both high $C_2$ selectivity and at the same time allow for high methane conversion.

A number of earlier studies suggested that a catalyst's basicity is the most important property influencing the conversion rate and selectivity [7,8,9,10,11,12]. Despite several proposed schemes for quantification, no universal method that would cover a useful prediction of properties for a large number of the known OCM catalysts has yet been suggested. More recent studies have suggested other quantitative parameters related to the OCM as potential descriptors, such as methane activation through homolytic dissociation when both ·$CH_3$ and H atoms are adsorbed on surface O atoms [13,15]. Despite some interesting observations, the heterolytic dissociation of methane has not been studied [14]. This may be the reason why found correlations lack the universality [15]. In other studies, the correlation between OCM performance and the properties of various materials was studied using big-data analysis with the application of artificial intelligence (AI) methodologies. In studies by Zavyalova [16], Suzuki [17] and Mine [18], the datasets were created based on published reports; Takahashi and coworkers used high-throughput experimental data that the authors generated in the respective groups [20,21,22,23,24]. Based on reaction conditions parameters, and properties of the chemical elements in studied catalytic materials, these authors found new regularities and successfully could design alternative OCM catalysts by applying supervised as well as unsupervised machine learning methods, data mining. Pirro et al. used kinetic modeling and basicity properties in a statistical analysis to obtain descriptor-property relationships [19], based on the prior knowledge in the field the results they obtained confirmed the state of knowledge.

In our work, we follow a different concept. Our design of new and alternative OCM materials is based on atomic-level insights that determine the physico-chemical factors responsible for the catalyst´s OCM performance. We found these factors by performing consistent measurements of the catalyst performance and through physico-chemical analysis and simulation of the adsorption properties of relevant molecules for the very same catalysts. Based on a study by Schmack et al. [25], who could show that well-performing OCM catalysts should contain chemical elements whose binary oxides are able to form thermodynamically stable carbonates, we designed an experimental study for oxide-supported carbonates [26]. The study provided a volcano-type correlation between the overall OCM performance ($C_2$-yield) and the formation energies of respective carbonates. We interpreted the identified correlations in terms of the site-isolation concept [26]. Based on this experimental observation, we developed a theory-guided workflow for the computational prediction of new OCM catalysts, which combines density functional theory (DFT) and AI methods (Scheme 1). Since the stability of carbonates is a useful descriptor of the OCM reaction [26], we developed a method based on *ab initio* thermodynamics that allowed us to predict the formation energies of carbonates. With that, we reformulated experimentally observed volcano-type dependence in terms of experimental vs. theoretically predicted correlation. Refining the dataset and using AI methods (data mining and symbolic regression) allowed us to identify peculiarities of materials with enhanced OCM performance. This we used in an *active* high-throughput screening process to search for new, promising OCM-catalyst candidates. Our screening process incorporated both general textbook knowledge and rules which result from application of AI. The iterative fueling of the AI-algorithm part provided more accurate predictions compared to the standard screening process. As a result of this procedure, we proposed a set of new OCM candidates, and validated our theoretical strategy experimentally.

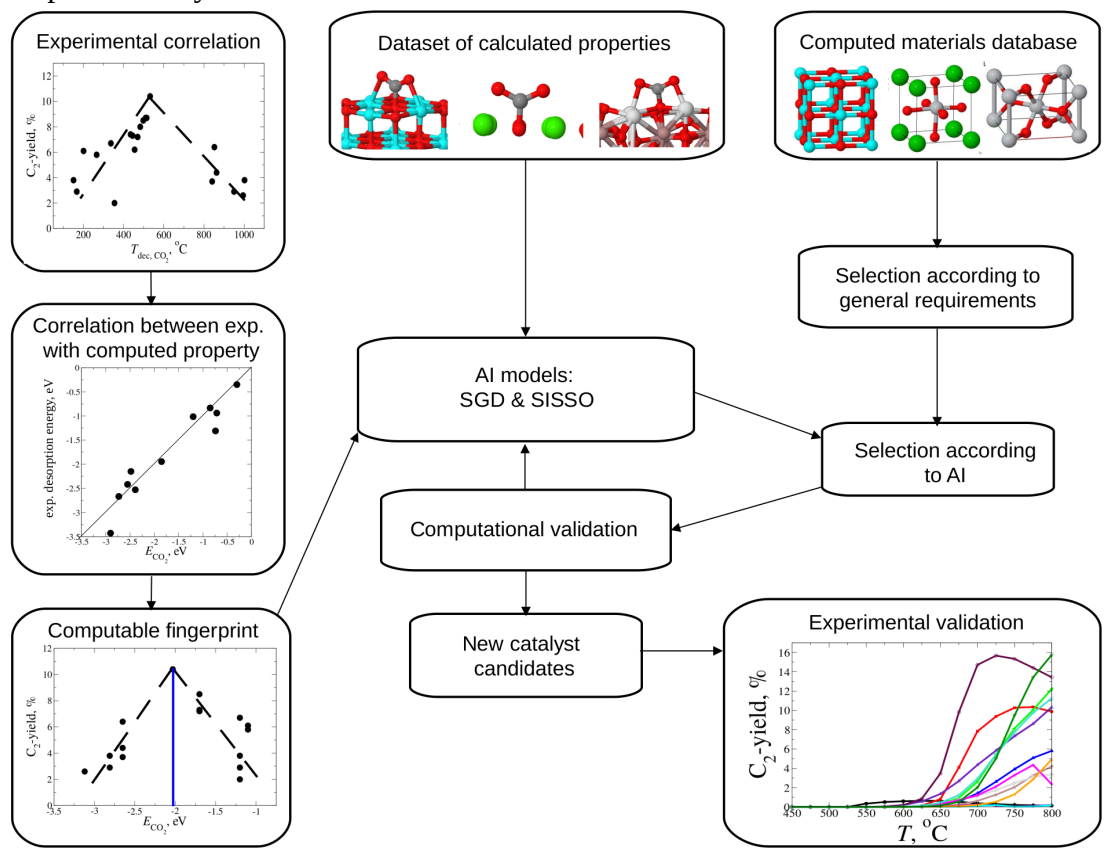

Scheme 1. The data-driven strategy for the search for new OCM catalysts.

## Results
### Formation energies of carbonates

Surface carbonates play decisive role in reactions catalyzed by oxides where carbon dioxide is either a reactant [28] or a byproduct [25,26], such as in OCM. In the last case, highly stable carbonates fully cover the surface, restricting the activation and conversion of reactants. The carbonates with low stability are completely decomposed at OCM conditions, so that the surface of a catalyst is bare, and through this bare surface also unselective sites are all exposed to the gas mixture. Carbonates with middle stability are present in tiny amounts under OCM conditions, and cover only highly reactive low-coordinated sites. This prevents the unselective oxidation of methane to carbon oxides; at the same time, most of the catalyst's surface area remains uncovered, providing sufficiently high conversion [26]. The effect of carbonate stability or decomposition temperature results in a volcano-type dependence of $C_2$-yield (ethane and ethylene) in the OCM reaction [26]. Thus, when designing new promising OCM catalysts, it is necessary to quantitatively assess the stability of carbonates formed on the corresponding oxide materials. Currently, the number of carbonates with reported decomposition energies determined experimentally or alternatively with decomposition temperatures is limited mainly to binary oxides of alkali, alkaline earth metals, some transition metals, and *f*-elements [25]. An alternative way to possibly design OCM catalysts would be to calculate the formation energies of carbonates, using first-principles methods. Such calculations imply differences in the total energies of gas-phase $CO_2$, oxides, and corresponding carbonates. Calculated structures for oxide materials are available in large quantities in many databases. In contrast, the number of calculated carbonates is much smaller. The search for new, previously unreported carbonate structures can be done by applying, for example, stochastic methods. However, this procedure is still quite computationally demanding.

Instead, we followed an alternative approach for the calculations of the formation energies of carbonates – one that does not involve an extensive search for their atomic structures. We proceed from the fact that carbonates are formed during the OCM reaction, first of all, on the surfaces of catalytic materials. Thus, the interaction of a $CO_2$ molecule with oxide surfaces might provide some information about a given carbonate's stability. Any direct match of calculated carbon dioxide adsorption energies with formation energies of carbonates is unlikely, because it is dependent on a surface termination and on the neighborhood of surface O atoms upon which $CO_2$ adsorbs – adsorption energies can be different. However, their averaging can provide very comprehensible results. We assume that the morphology of real surfaces is defined by the Boltzmann distribution of different surface terminations, dependent on their formation energies. Accordingly, the fragments of the lesser stable terminations should be more rarely observed on real materials, although they might be more chemically active, and might even trigger the whole catalytic reaction [30,14], so it is important to count them as well. With that, we considered the response of oxides upon carbon dioxide exposure ($E_{CO2}$) as an average adsorption energy sampled according to a Boltzmann distribution of corresponding surface terminations:

$$E_{CO_2} = \sum_i E_{ads,i}^{max} \frac{\exp\left(\frac{-E_{form,i}}{k_B T}\right)}{\sum_i \exp\left(\frac{-E_{form,i}}{k_B T}\right)}$$

(1)

where $E_{ads,i}^{max}$ – $CO_2$ adsorption energy on the *i*th surface termination, the maximal (most negative) among all possible values for different sites on a given surface; $E_{form,i}$ – formation energy of the *i*th surface termination; $k_B$ – the Boltzmann constant; and $T$ – temperature.

This distribution does not assume in general any degeneracies unless two surface terminations are enantiomers, which is, however, quite a rare phenomenon. In other words, all surface terminations, or at least adsorption sites delivering $E_{ads,i}$, have to be unique. As an illustration of non-unique surface terminations, one can consider (012) cut of cubic MgO. It consists of equal amounts of step and terrace fragments, and its formation energy is roughly equal to the average of the formation energies of (001) terrace and (110) stepped terminations (for more details, see Figure S1). To verify that all the considered surfaces are unique, we used many-body tensor representation (MBTR) [31] as the similarity measurement, implemented in DScribe package [32]. In MBTR the geometric structure is represented as a vector containing all the information about internal interatomic distances, angles, and torsion angles. Since a $CO_2$ molecule is activated on surface O atoms in the case of semiconductor oxides [27], the MBTR vectors of these atoms are obtained considering the first coordination shell of neighborhood atoms. The scalar products of such vectors for the different surface terminations provide information about their (dis)similarity. This makes possible to prevent multiple counts of the same adsorption sites in (1).

We validated this approach for the carbonates of Be, Ag, Pb, Li, Na, Mg, Ca, Sr, Ba, Zn, and La, for which standard formation enthalpies ($\Delta_f H^0_{298}$) have been reported [33,34]. For these carbonates, we considered the enthalpy-of-decomposition reaction ($\Delta_r H^0_{298}$): $Me_x(CO_3)_{y(s)} = Me_xO_{y(s)} + yCO_{2(g)}$ at $p(CO_2) = 1$ bar and $T = 298$ K. Since calculated $E_{CO2}$ values with our method only account for electron-electron, electron-nuclei, and nuclei-nuclei interactions, for validation we derived corresponding contributions from experimental $\Delta_f H^0_{298}$, neglecting the differences in vibrational contributions from solid phases (see the SI):

$$\Delta_r E_{exp} \approx \Delta_r H^0_{298} - \left( H^0_{298}(CO_2) - H^0_0(CO_2) \right) \quad (2)$$

in which $H^0_T(CO_2)$ is the enthalpy of carbon dioxide at temperature $T$. Corresponding values have been tabulated, for example, in JANAF [35]. The $E_{CO2}$ quantity is temperature-dependent. Here, we refer to the temperature at which experimental data have been reported (298 K), although, as shown below, this is not the only option. To calculate $E_{CO2}$ for each oxide, several unique surface terminations were considered, and the adsorption of $CO_2$ on them was calculated or taken from previous studies [27,36]. The number of considered terminations for Eq. (1) depends on the number of types of oxygen atoms in a given material (defined by coordination numbers and neighborhood atoms). In the case of binary oxides, we usually investigated between three and five terminations; for ternary and quaternary oxides, there were considered up to seven terminations.

In Figure 1 and Table S1, the quantities $\Delta_r E_{exp}$ and $E_{CO2}$ are compared. For most of the samples, we obtained a reasonable match, with a mean absolute deviation (MAD) 0.21 eV. This means that the proposed approach can be applied to estimate the formation energies of carbonates from a wide range of values.

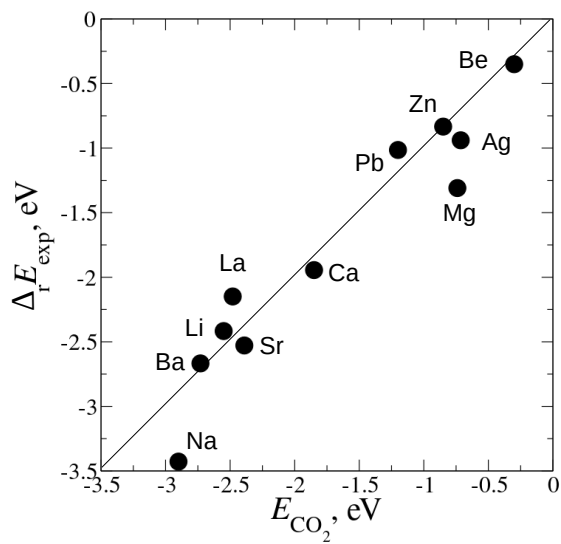

Figure 1. $\Delta_r E_{exp}$ derived from standard formation enthalpies of carbonates, oxides, and gas-phase $CO_2$ vs. calculated $E_{CO2}$ values for corresponding oxides.

We should also comment that our approach for prediction of the formation energies of carbonates is more accurate than Stern's model [37], which is simply the square root of cationic radii divided by effective nuclear charge. Stern's model is accurate enough for alkali and alkaline earth metals, but for some other elements it fails (Figure S2). For above mentioned carbonates, the MAD of Stern's linear fit is 0.6 eV – much less accurate than our approach.

**A computable fingerprint for the OCM reaction**

In the work of H. Wang [26] the fitting of OCM performance was done for the first and the last $CO_2$ desorption peaks obtained in thermogravimetric analysis (TGA) experiments with no reference about their origin. We observed that the lowest temperature peak had the best-pronounced volcano-type correlation. Most of the corresponding $CO_2$ decomposition temperatures coincided with water desorption peaks, implying hydroxycarbonates decomposition. However, the number of $CO_2$ desorption peaks for some studied materials in TGA experiments was up to three [38], indicating the availability of carbonates with different stability, in addition to hydroxycarbonates. Here we considered all the $CO_2$ desorption peaks obtained in TGA experiments, and tried to match them with calculated formation energies $E_{CO2}$ according to Eq. (1). For this reason, we carried out corresponding calculations both for binary oxides involved in experiments (oxides of Y, Al, Sm, Gd, Ce (IV), Mg, Ba, Sr) and for some ternary oxides, which potentially could be formed from two binary oxides/carbonates under increased temperatures in TGA experiments – $Ba_2MgO_3$, $BaMgO_2$, $Cs_2Sr_2O_3$, $Cs_3YO_3$, $Rb_3YO_3$, $RbYO_2$, $MgY_2O_4$, $Cs_2Ba_2O_3$, $Cs_2BaO_2$, $Cs_3AlO_3$, $CsAlO_2$, $Rb_3AlO_3$, $RbAlO_2$, $MgAl_2O_4$, $CsGdO_2$, $Mg_2Gd_2O_5$, and $MgCeO_3$.

For calculations of $E_{CO2}$, the temperatures in Eq. (2) were defined for each material individually. We proceed from the fact that the decomposition of carbonates in TGA experiments takes place at the temperatures when the change of Gibbs free energy for the process $CO_2@oxide_{(s)} = CO_{2(g)} +$ oxide$_{(s)}$ is zero. Assuming that the pressure of carbon dioxide in the TGA experiments was about 1 bar, the desorption energy is defined as:

$$\Delta_d E_{exp} \approx TS_{CO2}(T) - (H_{CO2}(T) - H_{CO2}(0))$$ (3)

where $T$ – decomposition temperatures of carbonates, $S_{CO2}$, $H_{CO2}$ – the entropy and enthalpy of gas-phase carbon dioxide at a given temperature (for more details, see the SI). In other words, the desorption energy $\Delta_d E_{exp}$ is the isobaric shift of $CO_2$ chemical potential at non-zero temperatures: $\mu_{CO2}(p,0) - \mu_{CO2}(p,T)$, since $\mu_{CO2} = H_{CO2} - TS_{CO2}$ per 1 mole. The temperature dependence of $E_{CO2}$ comes from the Boltzmann factor, which contains formation energies of surface terminations in the numerator. This means that less stable surface terminations can contribute to a larger degree at higher temperatures, influencing the final value of $E_{CO2}$. Therefore, for further analysis we take the $E_{CO2}$ energies calculated at the temperatures at which the $CO_2$ desorption becomes thermodynamically favorable, i.e., the condition $E_{CO2}(T) = \mu_{CO2}(0) - \mu_{CO2}(T)$ is fulfilled (Figure 2a).

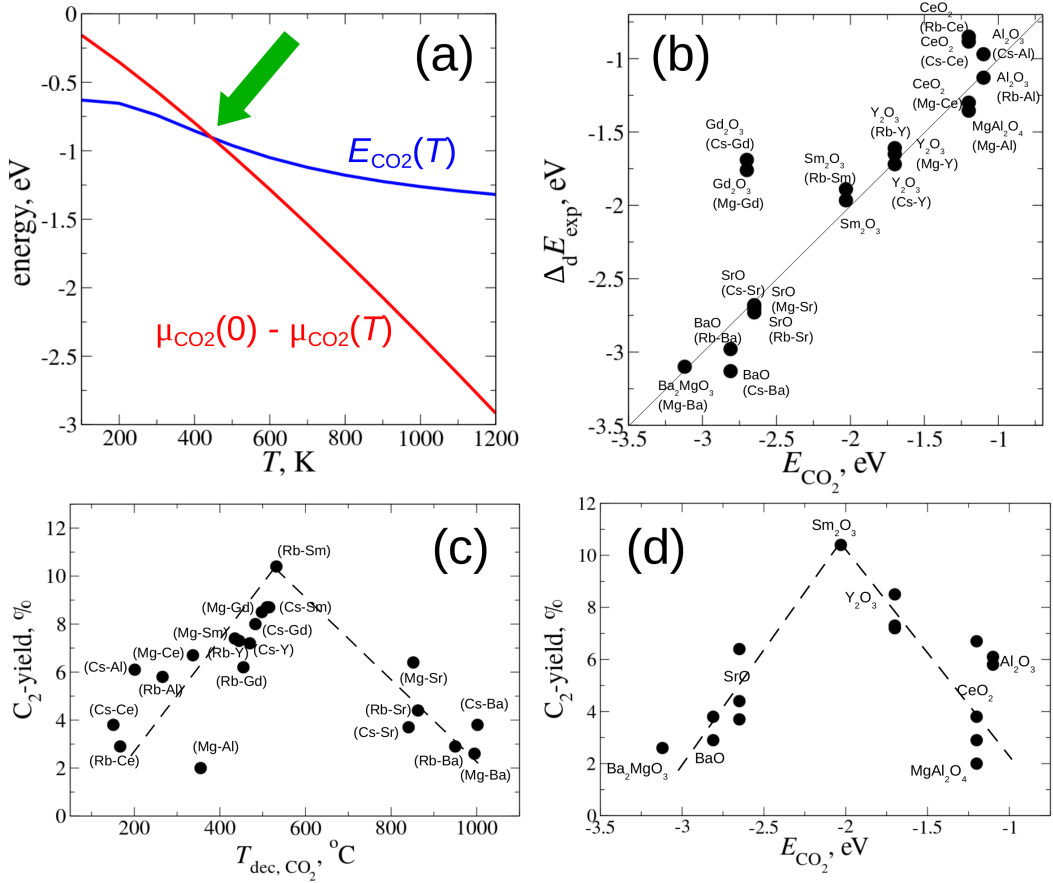

Figure 2. (a) Temperature dependence of $E_{CO2}(T)$ (blue) and $\mu_{CO2}(0) - \mu_{CO2}(T)$ (red) for MgO; the green arrow shows the cross-section point for which $E_{CO2}$ values have been calculated. (b) The correlation between $\Delta_d E_{exp}$ obtained from TGA experiments [26,38] and calculated $E_{CO2}$ energies; for each data point, the corresponding calculated material and the composition of a measured oxide (in parentheses) are shown (for experimental data, we follow the same abbreviations as in Ref. 26). (c) Experimentally observed volcano-type dependence of $C_2$-yield on the decomposition temperatures of carbonates (adapted from Ref. 26). (d) Dependence of $C_2$-yield on DFT calculated formation energies of carbonates, $E_{CO2}$.

Figure 2b depicts the correlation between desorption energies $\Delta_d E_{exp}$ derived from $T_{dec}$ and $E_{CO2}$. The calculated values are shown in Table S2. In most of the cases, the $E_{CO2}$ energies that we calculated sufficiently approximate the quantities obtained experimentally. This once again confirms the applicability of our approach for the calculation of the formation energies of carbonates. The variance of experimental decomposition temperatures for several compositions in which the same oxide is responsible for observed $T_{dec}$ (for example, $Y_2O_3$, $Sm_2O_3$) is caused by polarization effects of supporting materials [26]. The only exceptions to the general trend were the compositions with $Gd_2O_3$ as support. The calculated $E_{CO2}$ value of -2.77 eV is much more negative compared to those derived from experiments $\Delta_d E_{exp}$ (-1.69 eV for Cs-Gd-O and -1.76 eV for Mg-Gd-O). We also calculated $CO_2$ adsorption on $Gd_2O_3$ using a more advanced DFT hybrid method, HSE06 [39], which correctly describes strongly correlated $f$ systems [40]. The difference obtained with the HSE06 calculated value is relatively small, about 0.2 eV, suggesting that the reason is not in theory level. However, a detailed solution of this problem lies outside of the scope of the present study and does not compromise the main results.

The obtained correlation between $E_{CO2}$ and $\Delta_d E_{exp}$ allows us to replace experimentally measured decomposition temperatures of carbonates with their DFT-calculated counterparts. In such a way we reformulated the original OCM volcano-type plot (Figure 2c) in terms of the dependence of experimental $C_2$-yields vs. calculated $E_{CO2}$ energies (Figure 2d). The top of the OCM volcano in Figure 2c lies around samarium oxide, with a calculated formation energy of carbonate at -2.03 eV. We subsequently used this observation in the screening of materials.

**High-throughput screening**

The purpose of high-throughput screening campaign is to identify new oxide materials with potentially enhanced OCM performance. The descriptor of such performance is the ability to form carbonates with a decomposition energy close to the top of the volcano, i.e. -2 eV. Obviously, it is unfeasible to perform DFT calculations for each of the hundreds of thousands of oxide materials considering their surface terminations and corresponding $CO_2$ adsorption upon them. Therefore, we have applied AI methods to facilitate the screening campaign; these methods deliver relatively inexpensive models for the evaluation of the formation energies of carbonates.

**AI models**

To obtain the AI models, we extended our set of calculated formation energies of carbonates with binary and ternary oxides from Ref. 27. As a result, the initial training set consisted of 93 oxides with a range of $E_{CO2}$ values from -0.05 to -3.36 eV. Since the materials of our interest are those with $E_{CO2}$ values close to -2 eV, and consequently all others are out of our scope, we performed a subgroup discovery (SGD) to find the corresponding subgroup. For this reason, we rescaled our target property with respect to the top of volcano $E_{CO2,i}' = |E_{CO2,i} - -2|$. This allowed us to search the subgroups with quality function (5), minimizing the target property. The reason we chose to do so was that the volcano has nearly symmetric slopes around the top (Figure 2c). We used atomic features and the properties of bulk materials as the primary features for the SGD (Table

1).

Table 1. Primary features. All energy parameters are in eV, distances in Å, masses in atomic units.

| Feature | Description |
|---|---|
| $IP_{min,max,av,O}$ | ionization potential: minimal, maximal among all cations, averaged with respect to stoichiometry of cations, and for oxygen |
| $EA_{min,max,av,O}$ | electron affinity: minimal, maximal among all cations, averaged with respect to stoichiometry of cations, and for oxygen |
| $EN_{min,max,av,O}$ | Mulliken electronegativity: minimal, maximal among all cations, averaged with respect to stoichiometry of cations, and for oxygen |
| $PA_{min,max,av,O}$ | proton affinity: minimal, maximal among all cations, averaged with respect to stoichiometry of cations, and for oxygen |
| $r_{HOMO-1}$ | radius of HOMO-1: minimal, maximal among all cations, averaged with respect to stoichiometry of cations, and for oxygen |
| $r_{HOMO}$ | radius of HOMO: minimal, maximal among all cations, averaged with respect to stoichiometry of cations, and for oxygen |
| $r_{LUMO}$ | radius of LUMO: minimal, maximal among all cations, averaged with respect to stoichiometry of cations, and for oxygen |
| $\Delta$ | band gap |
| *wid* | width of projected O-2$p$ band |
| *skew* | skewness of projected O-2$p$ band |
| *kurt* | kurtosis of projected O-2$p$ band |
| *dens* | Density |
| *vol* | average atomic volume in a bulk material |
| $E_{at}$ | atomization energy per atom |
| $E_V$ | atomization energy per volume of a unit cell |
| $E_m$ | atomization energy per mass of a unit cell |

As we mentioned above, our assumption is that promising alternative OCM catalysts should be able to form carbonates with decomposition energies about 2 eV. However, the top of the volcano is not necessarily a sharp peak; it may be a plateau. Besides that, we also take into account that PBEsol-calculated $CO_2$ adsorption energies have an error of about 0.1 eV compared to high-level calculations and experimental data [27]. Thus, in order to estimate the range of the calculated $E_{CO2}$ values for materials with good OCM performance, we analyzed materials from our training set with respect to their reported catalytic measurements. Although the reaction conditions from previous studies were not exactly the same as in our work [26] (slightly higher and lower temperatures, He carrier gas instead of $N_2$, lower concentrations of methane and oxygen in gas mixtures, slightly different ratio $CH_4:O_2$ = 2:1 instead of 1.73:1), we nevertheless found enhanced OCM performance with $C_2$-yields ($Y_{C2}$) above 12% for several materials:
- $BaTiO_3$ ($E_{CO2}$ = -1.91 eV) with $Y_{C2}$ = 14.0% at 1023 K [43]
- $CaTiO_3$ ($E_{CO2}$ = -1.94 eV) with $Y_{C2}$ = 13.3% at 1073 K [44]
- $BaZrO_3$ ($E_{CO2}$ = -2.14 eV) with $Y_{C2}$ = 13.2% at 1093 K [45]
- $CaZrO_3$ ($E_{CO2}$ = -1.98 eV) with $Y_{C2}$ = 12.4% at 973 K [45]

- SrZrO$_3$ ($E_{CO2}$ = -2.19 eV) with Y$_{C2}$ = 14.3% at 1013 K [45]

In addition, we performed calculations for Li-MgO, which is one of the most popular catalysts for OCM [41]. Based on previous studies, we also considered several surfaces with Li dimers as substitutional defects of an Mg atom and with O vacancies stabilized with a pair of Li atoms [42] (for more details, see the SI). The $E_{CO2}$ energy calculated for Li-MgO was -2.04 eV. In sum, we concluded that promising OCM catalysts should possess calculated formation energies of carbonates of -2.0±0.2 eV, or up to 0.2 eV of shifted formation energies, $E_{CO2}'$.

The subgroup discovery performed for the discussed dataset resulted in a subgroup with a range of $E_{CO2}'$ values of 0.0-~0.6 eV. This means that not all materials from the subgroup can be expected to be promising. Therefore, we applied symbolic regression method SISSO to more accurately predict $E_{CO2}'$ values for samples within the identified subgroups.

**Selection**

The high-throughput screening was performed for calculated materials from the Open Quantum Materials Database (OQMD) [29], with 815654 structures up to the date of study. All materials available in the OQMD [29] passed through two types of criteria in a step-by-step manner – conditions imposed by general (textbook) knowledge, and conditions (models) obtained from SGD and SISSO (Figure 3a). These selection requirements are translated into the following selection steps:

- Since all materials in our training set are semiconductor oxides, selected materials should contain oxygen and have a non-zero band gap. Also, they should not contain radioactive elements, so that their potential usage could be safe for humans;
- Cations should have a maximal stable oxidation state, since the OCM reaction proceeds in an oxygen-rich atmosphere, implying that materials with chemical elements in a lower oxidation state can undergo oxidation and a corresponding loss of activity;
- At usual OCM reaction temperatures of about 600-800°C, phase transitions from metastable to more stable phases can occur. For this reason, we selected only the most thermodynamically stable polymorphs with the lowest calculated formation energies;
- Selected materials should fulfill conditions from selectors of subgroups obtained with SGD ($f_1 > a$, $f_2 \leq b$, etc.);
- All our calculations were made using PBEsol functional, whereas materials in the OQMD were calculated using other DFT methods. Thus, we performed single-point PBEsol calculations for all materials selected on previous steps for more accurate predictions. We then used the material features we obtained from these calculations in SISSO models to predict $E_{CO2}'$;
- For materials with a SISSO-predicted $E_{CO2}' < 0.2$ eV from single-point calculations, we performed full relaxation including unit-cell optimization in PBEsol, and the extracted features were used again to obtain more accurate SISSO predictions for $E_{CO2}'$.

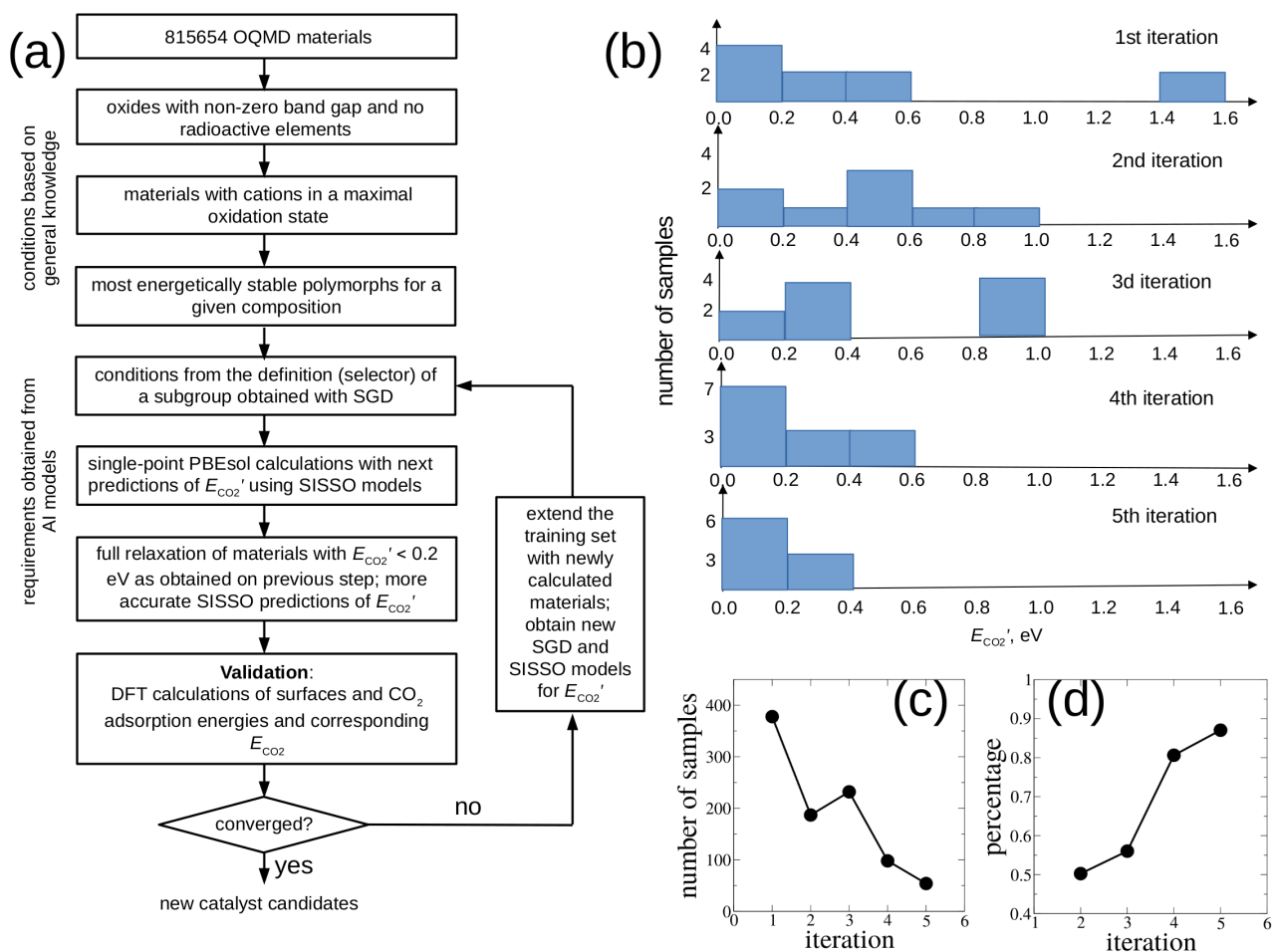

Figure 3. (a) The iterative high-throughput screening scheme; (b) the results of DFT validation at each iteration of the screening; (c) the number of samples according to criteria from a subgroup selector dependent on iteration; (d) the percentage of common samples in every $i$th iteration with samples obtained on all previous iterations.

Following this procedure we obtained about a hundred materials with SISSO predicted shifted formation energies of carbonates below 0.2 eV. Next, we performed the validation – DFT calculations of $E_{CO2}$ energies for ten randomly chosen materials with low predicted $E_{CO2}'$ values. The calculated shifts from the top of the volcano (-2 eV) for these materials are shown in Figure 3b, top. Four out of the ten materials possessed formation energies of carbonates within the expected -2.0±0.2 eV interval. Another four oxides deviated from the top of the volcano by up to 0.6 eV, which is acceptable because the range of $E_{CO2}'$ values in the obtained subgroup is exactly the same. However, two materials are obvious outliers, with $E_{CO2}' \approx 1.5$ eV, so according to our expectations, they are unlikely to be promising for OCM reactions. The reason why our prediction was not accurate enough was that the variety of materials in our training set was limited. As a result, the AI models, that are obtained based on an insufficient amount of data, failed for some oxide materials.

We solved this problem by performing *active* high-throughput screening, which included the iterative learning part during the AI stage. This means that after the screening of candidate materials applying AI models and validating these models by making DFT calculations of 8-13 promising candidates with small predicted $E_{CO2}'$ energies, we checked if the deviation of calculated $E_{CO2}'$ was larger than maximal $E_{CO2}'$ energy in the subgroup. If that was case, we retrained the SGD and SISSO models, and repeated the whole procedure (Figure 3a). After five iterations of this procedure, we observed that the shifted formation energies of carbonates for DFT-validated materials were less than or equal to 0.4 eV (Figure 3b) – within the same range as materials in the subgroup obtained from the training set that formed after the fourth iteration. With that we concluded that our iterative procedure had converged to a sufficient extent. Convergence is also supported by the observations that the number of candidate materials obtained according to the subgroup selectors decreased with each iteration (Figure 3c), and that the percentage of materials among them which had already appeared in any previous iteration approached 100% (Figure 3d). The observed changes were non-monotonic due to the stochastic algorithm in the SGD search. Therefore, at each iteration we dealt with more or less the same pool of materials, removing unpromising candidates and converging in such a way to a smaller, but more specific set of oxides with a narrow enough range of formation energies of carbonates. Compared to the non-iterative screening scheme our approach clearly provided more accurate predictions (Figure 3b).

In conclusion, our final subgroup had the selector ($PA_{av} \geq 9.46$ eV) AND ($E_V < -0.45$ eV/Å$^3$) AND ($r_{HOMO-1}^{min} \leq 0.88$ Å) AND ($r_{LUMO}^{av} \leq 1.75$ Å). The SISSO model for this subgroup is therefore the following:

$$E_{CO_2}' = 0.0312 + 0.1885 \left( \frac{wid \cdot r_{HOMO-1}^{max}}{r_{HOMO}^{O} + r_{HOMO-1}^{min}} - \frac{PA_O}{r_{HOMO-1}^{av}} \cdot \left( r_{HOMO-1}^{av} - r_{LUMO}^{min} \right) \right) \quad (4)$$

The fitting and cross-validation (CV) root-mean-square error (RMSE) of this model were 0.073 and 0.077 eV. Details about the cross-validation and the analysis of the models we obtained are described in the SI.

Next, we investigated the sintering characteristics of materials obtained in the high-throughput screening. According to Schmack et al. [25], sintering can be estimated using the Tammann temperatures ($T_T$) of binary oxides formed from cations which compose at least 50% of all the cations in a material. The requirement is that $T_T > 1073$ K. After adding this condition, we obtained 20 materials with a calculated $E_{CO2}$, and 11 materials with predicted $E_{CO2}'$ energies. These are listed in Table 4.

Table 4. Promising oxide materials for the OCM reaction with corresponding space groups, indices in ICSD, and formation energies of carbonates.

| Formula | Space group | ICSD ID | Calculated $E_{CO2}$ | SISSO-predicted $E_{CO2}'$[a] | Iteration number |
|---|---|---|---|---|---|
| $Sr_3B_2O_6$ | R-3c | 93395 | -2.07 | | 1 |
| $Ba_5Nb_4O_{15}$ | P-3m1 | 157477 | -2.13 | | 1 |
| $Ba_3MgTa_2O_9$ | P-3m1 | 163553 | -2.06 | | 1 |
| $Ba_2TaGaO_6$ | I4/mmm | | -2.12 | | 1 |
| $LuNbO_4$ | C2/c | 109182 | -1.98 | | 2 |
| $Y_2GeO_5$ | C2/c | 260425 | -1.87 | | 2 |
| $Nd_2Zr_2O_7$ | Fd-3m | 62793 | -2.14 | | 3 |
| $Nd_2TiO_5$ | Pnma | 2898 | -2.03 | | 3 |
| $Ba_2SnO_4$ | Pccn | 27115 | -1.85 | | 4 |
| $CaHfO_3$ | Pnma | | -1.89 | | 4 |
| $BaHfO_3$ | Pm-3m | | -2.10 | | 4 |
| $Ba_2CaTeO_6$ | I4/m | | -1.93 | | 4 |
| $Ca_2HoNbO_6$ | P21/c | 86297 | -2.00 | | 4 |
| $Ba_2SrIn_2O_6$ | I4/mmm | 66018 | -2.16 | | 4 |
| Ca$_2$LuNbO$_6$ | P2$_1$/c | 86299 | -1.99 | | 5 |
| Ca$_2$ErNbO$_6$ | P2$_1$/c | 86298 | -2.15 | | 5 |
| La$_2$SO$_6$ | C2/c | 66823 | -1.87 | | 5 |
| Sr$_2$TiO$_4$ | I4/mmm | 20293 | -1.83 | | 5 |
| NaLa$_2$TaO$_6$ | P2$_1$/c | 159206 | -1.94 | | 5 |
| Sr$_4$Nb$_2$O$_9$ | P-1 | 79217 | -2.15 | | 5 |
| Ca$_2$ZrO$_4$ | Pbam | | | 0.06 | 5 |
| Ba$_3$Ta$_2$NiO$_9$ | P-3m1 | 240281 | | 0.15 | 5 |
| Ca$_2$NbAlO$_6$ | P2$_1$/c | 172327 | | 0.18 | 5 |
| BaCa$_2$V$_2$O$_8$ | R-3 | | | 0.06 | 5 |
| BaNd$_2$Sc$_2$O$_7$ | P4$_2$/mnm | 167601 | | 0.05 | 5 |
| BaSm$_2$Sc$_2$O$_7$ | P4$_2$/mnm | 167602 | | 0.06 | 5 |
| BaGd$_2$Sc$_2$O$_7$ | P4$_2$/mnm | 167604 | | 0.07 | 5 |
| Ba$_2$DyTaO$_6$ | I4/m | | | 0.18 | 5 |
| La$_3$BWO$_9$ | P6$_3$ | 39809 | | 0.19 | 5 |
| Ba$_3$V$_2$O$_8$ | R-3m | 418460 | | 0.05 | 5 |
| La$_3$FeO$_6$ | Cmc2$_1$ | 421426 | | 0.05 | 5 |

[a] SISSO-predicted formation energies are shown only for materials obtained in the last iteration

We should mention that some of the predicted materials shown in Table 4 have been already studied in the OCM reaction. Sr$_2$TiO$_4$ has been reported in several studies [18], and at 800°C the C$_2$-yield was obtained up to 11.7% [46], whereas another study at 700°C found a yield of 18.6% [47]. A composition with Nd:Zr = 1:1 (similar to Nd$_2$Zr$_2$O$_7$ obtained during our screening) was prepared with sol-gel method and demonstrated the yield 3.5% at 700°C [48]. As we show in Figure 4a for predicted Sr$_3$B$_2$O$_6$ and the reference material Mn-Na$_2$WO$_4$/SiO$_2$, the C$_2$-yields at the same temperature are 2.6% and 2.0% respectively, and they rapidly increase as the temperature increases. Similar rapid yield growth was observed for the best catalyst in the original volcano-type plot, Rb-Sm oxide (Figure 2c), for which the C$_2$-yield is about 2.3% at 700°C [26]. Therefore, we expect that similar behavior can be observed for Nd$_2$Zr$_2$O$_7$ at $T > 700$°C. In conclusion, these studies either confirm or at least do not contradict our predictions.

**Experimental validation**

For experimental validation, we considered a set of oxide materials with a wide range of calculated formation energies of corresponding carbonates from -2.43 to -0.44 eV, including several materials with $E_{CO2}$ values close to -2 eV (Table 5). On the one hand, the reason to investigate a broad set was to verify that the general volcano trend, reformulated in terms of DFT-calculated $E_{CO2}$, is robust; on the other hand, concentrating on a relatively narrow subset of materials reveals issues with their synthesis. We measured OCM performance for all synthesized materials (Table S5). Since some target compositions were found to be unstable under reaction conditions (YInO$_3$), or the materials were polycrystalline and contained side phases (see the SI), we refer here to the main phase of a given material observed with XRD analysis after OCM reaction. In addition to that, we also included two reference catalysts with well-known OCM performance – CaO [6] and Mn-Na$_2$WO$_4$/SiO$_2$ [49]. The results of OCM measurements for these catalysts performed at 800°C are shown in Table 5.

Table 5. Materials for which experimental OCM study was performed with corresponding BET surface area, DFT formation energies of carbonates ($E_{CO2}$), and $C_2$-yields.

| Composition | Surface area (BET), $m^2/g$ | Calculated $E_{CO2}$, eV | $C_2$-yield at 800°C, % |
|---|---|---|---|
| $Sr_3B_2O_6$ | 2.5 | -2.07 | 12.2 |
| $Ba_2GaTaO_6$ | 13.2 | -2.12 | 13.4 |
| $Mn-Na_2WO_4/SiO_2$ (reference) | | -1.86[a] | 15.8 |
| CaO (reference) | 5.7 | -2.28 | 9.9 |
| $RbAlO_2$ | | -2.38 | 10.3 |
| $Ba_2MgMoO_6$ | 1.0 | -2.43 | 11.2 |
| $Ca_2AlTaO_6$ | 22.7 | -1.71 | 2.35 |
| $ScAlO_3$ | 20.9 | -1.57 | 4.2 |
| $ZrO_2$ | 5.1 | -1.58 | 3.4 |
| $Ba_2SiO_4$ | 5.7 | -1.43 | 4.9 |
| $CaNb_2O_6$ | 16.5 | -1.42 | 5.8 |
| $In_2O_3$[b] | 26.2 | -1.23 | 0.9 |
| $SrCrO_4$ | 6.9 | -0.44 | 0.2 |

[a] – carbonate-formation energy estimated based on $CO_2$ desorption temperature reported in Ref. 38;
[b] – the target composition $YInO_3$ was not observed in XRD data after the OCM reaction.

Figure 4 shows the relationship of $C_2$-yields obtained at 800°C on the calculated formation energies of carbonates, as well as the maxima yields for some catalysts obtained at lower temperatures. The original volcano trend shown in Figure 2c is clearly reproduced. Moreover, the catalysts with enhanced catalytic performance lie in the aforementioned range of carbonate-formation energies -2.0±0.2 eV. This validates our approach for predicting the AI-assisted carbonate-formation energies as a descriptor of OCM performance.

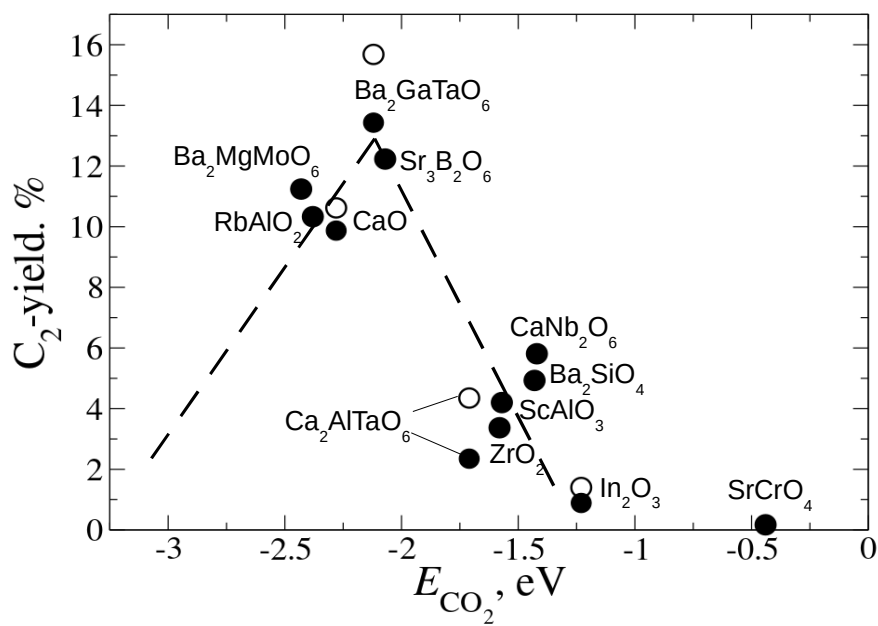

Figure 4. $C_2$-yields obtained at 800°C (filled circles) and maximal obtained yields at lower temperatures (empty circles) in relation to DFT-calculated formation energies of carbonates ($E_{CO2}$) for corresponding materials.

The reported maximal $C_2$-yield in Ref. 26 for a samarium-oxide-based system is 10.4%. As one can see in Figure 4, several materials had yields above 10%. A subgroup of some of them (CaO, $RbAlO_2$, $Ba_2MgMoO_6$) is located at carbonate-formation energies -2.4 – -2.3 eV – slightly below the top of the volcano. However, for the materials with closest to -2 eV carbonate-formation energies the highest $C_2$-yields are obtained: 12.2% for $Sr_3B_2O_6$ and 13.4% for $Ba_2GaTaO_6$ at 800°C. To the best of our knowledge, neither of these oxides have ever been reported in the context of OCM catalysis. Interestingly, similarly to CaO, the performance of $Ba_2GaTaO_6$ at lower temperatures is significantly higher compared to the performance of a standard $Mn-Na_2WO_4/SiO_2$ catalyst with the yield of 15.7% at 725°C (Figure S5). This observation shows that the condition of -2.0±0.2 eV for the formation energy of carbonates is also relevant for catalysts being active at lower temperature in the OCM reaction. We should emphasize here that these experiments were performed with the aim of validating our theoretical concept and workflow, and not to obtain higher $C_2$-yields with new catalyst candidates. Besides the carbonates' stability there are other certainly yet to be discovered descriptors of OCM performance, since materials with similar $E_{CO2}$ values provided variable yields. This will be a topic for future studies. Moreover, the synthesis methods of the AI proposed catalysts, reaction conditions and feed compositions can all be further modified, thus clearing a wide field for alternative studies.

**Conclusion**

We developed a data-driven strategy for the search for new catalytic materials and applied it to the OCM reaction. Our strategy combined consistent experimental data, *ab initio* thermodynamics calculations, and the application of AI methods. This strategy was based on our recent study [26]: while performing consistent OCM experiments we observed that carbonate species isolate unselective sites. This observation is expressed in a volcano-type dependence of $C_2$-yield on carbonate stability, so that only oxide-forming carbonates with moderate levels of stability show enhanced OCM performance. To search for new catalytic materials, we thus developed an approach to inexpensively calculate carbonate-formation energies. In this approach, the DFT-calculated $CO_2$ adsorption energies on unique surface terminations of a given oxide can be sampled according to the Boltzmann distribution of formation energies of corresponding facets. When we validated the calculated results with respect to experimentally measured formation enthalpies of carbonates, we found a mean deviation error of our approach of about 0.2 eV. This method thus allowed us to reformulate the initial experimentally observed volcano-type dependence in terms of experimental vs. calculated correlation. We then used this to conduct the high-throughput screening of hundreds of thousands of computed materials available in the OQMD [29]. Our screening procedure included both general, "textbook" requirements (the stability of materials under reaction conditions, the absence of radioactive chemical elements, etc.) as well as requirements obtained from AI analysis (SGD data mining and SISSO symbolic regression). The AI steps of the screening incorporated "active learning" – in each iteration, DFT calculations of carbonate-formation energies were done for about ten predicted materials. In the case prediction accuracy was not acceptable, the AI models were retrained using the extended training set incorporating the newly calculated materials, and the screening procedure was repeated. We demonstrated that *active* screening delivered more accurate predictions compared to the traditional screening procedure. As a result, we predicted that about thirty previously unreported materials could be promising catalysts for OCM reaction. We then synthesized some materials that had a relatively wide range of calculated carbonate-formation energies, and measured their OCM performance. The resulting $C_2$-yields closely followed the same volcano-type relationship that formed the basis of the material search strategy, thus validating our strategy. The best catalyst candidates, $Sr_3B_2O_6$ and $Ba_2GaTaO_6$, reach maximum $C_2$-yields comparable to the well-known OCM catalyst Mn-$Na_2WO_4$/$SiO_2$, and outperform this system at lower temperatures ($Ba_2GaTaO_6$).

**Methods**
**Computational details**

All DFT calculations were done using the PBEsol exchange-correlation functional [50]. We validated this approach in our previous work [27] in which we demonstrated that $CO_2$ adsorption energies calculated in PBEsol, among other GGA functionals, most closely match experiment-derived data and adsorption energies obtained using high-level methods. Most of the DFT calculations were performed using all-electron FHI-aims code [51] with numerical atomic orbitals. We used 'tight' basis sets according to the code's internal classification [51]. Zero-order regular approximation (ZORA) was used to account for relativistic effects [52]. Since standard GGA methods do not correctly treat systems containing *f* states due to overdelocalization, we used a pseudopotential approach as implemented in Quantum-Espresso code for calculations of corresponding systems [53]. In these cases, *f* states are a part of the pseudopotential and thus are not explicitly treated in SCF procedure. The valence orbitals are described with projector-augmented waves (PAW) [54]. As shown, this approach turned out to be reasonable for the calculations of $CO_2$ adsorption. We used pseudopotentials with a scalar relativistic treatment of core levels. The cutoff energy of PAW was set to 517 eV.

The calculations of adsorption were done using a slab-periodic model. For all materials in the study, we generated symmetric slabs with a thickness of at least 10 Å. A carbon dioxide molecule was always placed on one side of a slab, so that a dipole correction was used. The supercells were built with lattice vectors of a length above 9 Å to simulate low coverage. The vacuum gap between slabs was set to about 190 Å in the calculations in FHI-aims, and 20 Å in the case of Quantum-Espresso. The *k* grids for slabs were set for each surface individually, based on the converged grids for the corresponding bulk unit cells. The surfaces were then built by applying the Python ASE package [55]. The majority of the surfaces in this study were non-polar. Nonetheless, we also investigated some polar surfaces – particularly those with known reconstructions from the literature.

In some cases, these reconstructions were built artificially for supercells, removing atoms in the first few atomic layers. For relaxation jobs, we used the BFGS algorithm allowing the relaxation of all atoms in a slab. The accuracy of force calculations was $10^{-4}$ eV/Å. The search for local minima on a potential energy surface was carried out until maximal atomic forces did not exceed $10^{-2}$ eV/Å.

The adsorption energies were calculated as the total energy difference between a slab with adsorbed $CO_2$ and a free surface slab and a gas-phase carbon dioxide molecule.

To search for statistically exceptional subgroups, we applied a data-mining technique – subgroup discovery (SGD) [56,57]. This method was recently used in the field of heterogeneous catalysis to identify catalyst genes responsible for the activation of carbon dioxide [27] and to explain trends in hydrogen activation on single-atom alloys [58]. In SGD, the subgroups are defined with a set of binary selectors, which are inequalities of the type $f_1 > a$, $f_2 \leq b$, where $f_1$, $f_2$ are features of samples, and $a$ and $b$ are thresholds. These thresholds for each feature are identified as borders between neighboring clusters of data, using the $k$-means clustering algorithm to identify them. In our investigation, we used $k = 15$, which means that for each feature, 28 binary statements are generated. The algorithm searches for the subgroups with the unique distribution of the given target property within the whole dataset. This uniqueness is assessed with a quality function. In this study, we used the quality function developed by Boley et al. [59]:

$$F(X) = \frac{s(X)}{s(Y)} \left( \frac{med(Y) - med(X)}{med(Y) - min(Y)} \right) \left( 1 - \frac{amd(X)}{amd(Y)} \right) \quad (5)$$

with $X$ – subgroup, $Y$ – whole sampling, $s$ – size, $min$ and $med$ – minimum and median values of a target property, $amd$ – absolute average deviation of target property values around the median. Initial subgroups are generated as a conjunction of binary statements: seeds. The search for the most exceptional subgroups is done with a Monte Carlo algorithm in which a pruning of binary statements is randomly done and a quality function is calculated for each seed. We considered 100,000 seeds, and extracted the subgroups with the highest values of the quality function for analysis. SGD was performed using RealKD (https://bitbucket.org/realKD/realkd/).

The sure-independence screening and sparcifying operator (SISSO) method [60] was employed to obtain the models needed for quantitative predictions. This is compressed-sensing symbolic regression with $l_0$ regularization, which provides the models as superpositions of non-linear combinations of input features: $P = c_0 + \sum_{i=1}^{D} c_i d_i$, where $P$ – the target property, $c$ – linear coefficients, $d$ – non-linear combinations of primary features (complex features) and $D$ – the dimensionality of the descriptor. The compressed-sensing component of SISSO selects a $D$-dimensional combination of complex features with the lowest root-mean-square error (RMSE) out of billions of candidates. These candidates are all possible combinations of primary features, combined with a set of mathematical operators. Here, we used the following operators: +, -, *, /, exp(), ln(), powers -1, 2, 3, 1/2, 1/3, and absolute difference. The complexity of each complex feature Φ is a hyperparameter, equal to the maximum number of primary features in $d_i$. Thus, two hyperparameters, $D$ and Φ, define the predictive ability of a SISSO model. In order to define the optimal values of these hyperparameters, we applied leave-one-out cross validation (CV). During this procedure, the sample, consisting of $N$ samples, is split into $N$ subsets of $N$-1 size. For each subset, an individual SISSO model is obtained. Afterwards, these models are used to predict target property values for hold-out samples. The optimal values of $D$ and Φ are the ones which deliver the lowest overall error (CV-RMSE) of such predictions. In addition to complexity and dimensionality, prediction accuracy depends on the selection of primary features. To search for the optimal set of features, we used a genetic algorithm. The details of this algorithm will be published elsewhere;

here we just briefly describe what it is. A genetic algorithm is a technique that allows the user to find a set of primary features with the lowest CV-RMSE out of all possible ones. The algorithm works sequentially, and at each step new candidate sets of primary features are generated from two parent sets via crossover (50% of features are taken from each parent) with optional mutation (when a certain feature can be replaced by another one). The selection of each parent depends on a fitness function, which is in our case inversely proportional to CV-RMSE.

**References**
1. G.E. Keller, M.M. Bhasin. Synthesis of Ethylene via Oxidative Coupling of Methane: I. Determination of Active Catalysts. *J. Catal*. **1982**, *73*, 9.
2. V.I. Alexiadis, M. Chaar, A. van Veen, M. Muhler, J.W. Thybaut, G.B. Marin. *Appl. Catal. B Environ*. **2016**, *199*, 252.
3. Y. Gambo, A.A. Jalil, S. Triwahyono, A.A. Abdulrasheed. *J. Ind. Eng. Chem*. **2018**, *59*, 218.
4. R. Schlögl, *Angew. Chem. Int. Ed*. **2015**, *54*, 3465.
5. Y.S. Su, J.Y. Ying, W.H. Green Jr. *J. Catal*. **2003**, *218*, 321.
6. L. Thum, M. Rudolph, R. Schomäcker, Y. Wang, A. Tarasov, A. Trunschke, R. Schlögl. *J. Phys. Chem. C*. **2019**, *123*, 8018.
7. J.A.S.P. Carreiro, M. Baerns. *J. Catal*. **1989**, *117*, 396.
8. V.R. Choudhary, S.A.R. Mulla, B.S. Uphade. *Fuel*. **1999**, *78*, 427.
9. V.R. Choudhary, V.H. Rane, R.V. Gadre. *J. Catal*. **1994**, 145, 300.
10. V.R. Choudhary, V.H. Rane, M.Y. Pandit. *J. Chem. Tech. & Biotech*. **1997**, *68*, 177.
11. A.M. Maitra, I. Campbell, R.J. Tyler. *Appl. Catal. A: Gener*. **1992**, *85*, 27.
12. V.D. Sokolovskii, S.M. Aliev, O.V. Buyevskaya, A.A. Davydov. *Catal. Today*. **1989**, *4*, 293.
13. G. Kumar, S.L.J. Lau, M.D. Krcha, M.J. Janik. *ACS Catal*. **2016**, *6*, 1812.
14. A. Mazheika, S.V. Levchenko. *J. Phys. Chem. C*. **2016**, *120*, 26934.
15. S. Lim, J.-W. Choi, D.J. Suh, K.H. Song, H.C. Ham, J.-M. Ha. *J. Catal*. **2019**, *375*, 478.
16. U. Zavyalova, M. Holena, R. Schlogl and M. Baerns. *ChemCatChem*, **2011**, *3*, 1935.
17. K. Suzuki, T. Toyao, Z. Maeno, S. Takakusagi, K.-I. Shimizu, I. Takigawa. *ChemCatChem* **2019**, *11*, 1.
18. S. Mine, M. Toyao, T. Yamaguchi, T. Toyao, Z. Maeno, S.M.A.H. Siddiki, S. Takakusagi, K.-I. Shimizu, I. Takigawa. *ChemCatChem* **2021**, *13*, 3636.
19. L. Pirro, P.S.F. Mendes, S. Paret, B.D. Vandegehuchte, G.B. Marin, J.W. Thybaut. *Catal. Sci. & Technol.*, **2019**, *9*, 3109.
20. J. Ohyama, T. Kinoshita, E. Funada, H. Yoshida, M. Machida, S. Nishimura, T. Uno, J. Fujima, I. Miyazato, L. Takahashi, K. Takahashi. *Catal. Sci. & Technol*. **2021**, *11*, 524-530.
21. K. Takahashi, L. Takahashi, T.N. Nguyen, A. Thakur, T. Taniike. *J. Phys. Chem. Lett*. **2020**, *11*, 6819.
22. T.N. Nguyen, S. Nakanowatari, T.P. Nhat Tran, A. Thakur, L. Takahashi, K. Takahashi, T. Taniike. *ACS Catal*. **2021**, *11*, 1797.
23. L. Takahashi, T.N. Nguyen, S. Nakanowatari, A. Fujiwara, T. Taniike, K. Takahashi. *Chem. Sci.*, **2021**, *12*, 12546.
24. K. Sugiyama, T.N. Nguyen, S. Nakanowatari, I. Miyazato, T. Taniike, K. Takahashi. *ChemCatChem*. **2021**, *13*, 952.
25. R. Schmack, A. Friedrich, E.V. Kondratenko, J. Polte, A. Werwatz, R. Kraehnert. *Nat. Commun*. **2019**, *10*, 441.
26. H. Wang, R. Schmack, S. Sokolov, E. Kondratenko, A. Mazheika, R. Kraehnert. *ACS Catal*. **2022**, *12*, 9325.

27. A. Mazheika, Y. Wang, R. Valero, L.M. Ghiringhelli, F. Viñes, F. Illas, S.V. Levchenko, M. Scheffler. *Nat. Commun.* **2022**, *13*, 419.
28. M.M. Millet, G. Algara-Siller, S. Wrabetz, A. Mazheika, F. Girgsdies, D. Teschner, F. Seitz, A. Tarasov, S.V. Levchenko, R. Schlögl, E. Frei. *JACS*, **2019**, *141*, 2451.
29. J.E. Saal, S. Kirklin, M. Aykol, B. Meredig, C. Wolverton. *JOM* **2013**, *65*, 1501.
30. P. Schwach, N. Hamilton, M. Eichelbaum, L. Thum, T. Lunkenbein, R. Schlögl, A. Trunschke. *J. Catal.* **2015**, *329*, 574.
31. H. Huo, M. Rupp. *Mach. Learn.: Sci. Technol.* **2022**, DOI: 10.1088/2632-2153/aca005.
32. L. Himanen, M.O.J. Jäger, E.V. Morooka, F.F. Canova, Y.S. Ranawat, D.Z. Gao, P. Rinke, A.S. Foster. *Comp. Phys. Comm.* **2020**, *247*, 106949.
33. K.H. Stern, E.L. Weise. High-Temperature Properties and Decomposition of Inorganic Salts. Part 2, Carbonates. NSRDS-NBS 30. **1969**.
34. A. Olafsen Sjåstad, H. Fjellvåg, K.B. Helean, A. Navrotsky. *Therm. Acta*. **2012**, *550*, 76.
35. D.R. Stull, H. Prophet. JANAF thermochemical tables. *J. Phys. Chem.* **1974**, *78*, 2496.
36. Q.-L. Tang, Q.-H. Luo. *J. Phys. Chem. C*. **2013**, *117*, 22954.
37. K.H. Stern. *J. Chem. Ed.* **1969**, *46*, 645.
38. H. Wang. PhD thesis. Technische Universität Berlin, **2018**.
39. A.V. Krukau, O.A. Vydrov, A.F. Izmaylov, G.E. Scuseria. *J. Chem. Phys.* **2006**, *125*, 224106.
40. R. Gillen, S.J. Clark, J. Robertson. *Phys. Rev. B*. **2013**, *87*, 125116.
41. S. Arndt, G. Laugel, S.V. Levchenko, R. Horn, M. Baerns, M. Scheffler, R. Schlögl, R. Schomäcker. *Catal. Rev.: Sci. & Eng.* **2011**, *53*, 424.
42. P. Myrach, N. Nilius, S.V. Levchenko, A. Gonchar, T. Risse, K.-P. Dinse, L.A. Boatner, W. Frandsen, R. Horn, H.-J. Freund, R. Schlögl, M. Scheffler. *ChemCatChem* **2010**, *2*, 854-862.
43. M. Teymouri, E. Bagherzadeh, C. Petit, J. L. Rehspringer, S. Libs, A. Kiennemann. *J. Mater. Sci.* **1995,** *30*, 3005.
44. C.Y. Yu, W.Z. Li, G.A. Martin, C. Mirodatos. *Appl. Catal. A: General.* **1997**, *158*, 201.
45. N. Matsuda, K. Ohyachi, I. Matsuura. *Chem. Express*. **1990**, *5*, 533.
46. Y.A. Ivanova, E.F. Sutormina, N.A. Rudina, A.V. Nartova, L.A. Isupova. *Catal. Comm.* **2018**, *117*, 43.
47. D.V. Ivanov, L.A. Isupova, E. Y. Gerasimov, L.S. Dovlitova, T.S. Glazneva, I.P. Prosvirin. *Appl. Catal. A: Gen.* **2014**, *485*, 10.
48. C. Petit, A. Kaddouri, S. Libs, A. Kiennemann, J. Rehspringer, P. Poix. *J. Catal.* **1993**, *140*, 328.
49. X.P. Fang, S.B. Li, J.Z. Lin, Y.L. Chu, *J. Mol. Catal. (China).* **1992**, *6*, 427.
50. J.P. Perdew, A. Ruzsinszky, G.I. Csonka, O.A. Vydrov, G.E. Scuseria, L.A. Constantin, X. Zhou, K. Burke. *Phys. Rev. Lett.,* **2008**, *100*, 136406.
51. V. Blum, R. Gehrke, F. Hanke, P. Havu, V. Havu, X. Ren, K. Reuter, M. Scheffler. *Comp. Phys. Comm.* **2009**, *180*, 2175.
52. E.V. Lenthe, E.J. Baerends, J.G. Snijders. *J. Chem. Phys.* **1993**, *99*, 4597.
53. P. Giannozzi, S. Baroni, N. Bonini, M. Calandra, R. Car, C. Cavazzoni, D. Ceresoli, G.L. Chiarotti, M. Cococcioni, I. Dabo, A. Dal Corso, S. de Gironcoli, S. Fabris, G .Fratesi, R. Gebauer, U. Gerstmann, C. Gougoussis, A. Kokalj, M. Lazzeri, L. Martin-Samos, N. Marzari, F. Mauri, R. Mazzarello, S. Paolini, A. Pasquarello, L. Paulatto, C. Sbraccia, S. Scandolo, G. Sclauzero, A.P Seitsonen, A. Smogunov, Paolo Umari, R.M. Wentzcovitch. *J. Phys.: Condens. Matter* **2009**, *21*, 395502.
54. P.E. Blöchl. *Phys. Rev. B*. **1994**, *50*, 17953.
55. A.H. Larsen, J.J. Mortensen, J. Blomqvist, I.E. Castelli, R. Christensen, M. Dułak, J. Friis, M.N. Groves, B. Hammer, C. Hargus, E.D. Hermes, P.C. Jennings, P.B. Jensen, J .Kermode, J.R. Kitchin, E.L. Kolsbjerg, J .Kubal, K. Kaasbjerg, S. Lysgaard, J. Bergmann Maronsson, T .Maxson, T. Olsen,

L. Pastewka, A. Peterson, C. Rostgaard, J. Schiøtz, O. Schütt, M. Strange, K.S. Thygesen, T. Vegge, L. Vilhelmsen, M. Walter, Z. Zeng, K.W. Jacobsen. *J. Phys.: Condens. Matter.* **2017**, *29*, 273002.

56. S. Wrobel. *European Symposium on Principles of Data Mining and Knowledge Discovery* (*Springer*). **1997**, 78.

57. M. Atzmueller. *Data Min. Knowl. Disc*. **2015**, *5*, 35.

58. Z.-K. Han, D. Sarker, R. Ouyang, A. Mazheika, Y. Gao, S.V. Levchenko. *Nat. Comm*. **2021**, *12*, 1833.

59. M. Boley, B. Goldsmith, L.M. Ghiringhelli, J. Vreeken. *Data Min. Knowl. Discov*. **2017**, *31*, 1391.

60. R. Ouyang, S. Curtarolo, E. Ahmetcik, M. Scheffler, L.M. Ghiringhelli. *Phys. Rev. M*. **2018**, *2*, 083802.